\documentclass[conference]{IEEEtran}
\IEEEoverridecommandlockouts
\usepackage{hyperref}
\usepackage{cite}
\usepackage{CJKutf8}
\usepackage{amsmath,amssymb,amsfonts}
\usepackage{algorithmic}
\usepackage{graphicx}
\usepackage{textcomp}
\usepackage{xcolor}
\usepackage{multirow}
\usepackage{bbding}
\usepackage{siunitx}

\def\BibTeX{{\rm B\kern-.05em{\sc i\kern-.025em b}\kern-.08em
    T\kern-.1667em\lower.7ex\hbox{E}\kern-.125emX}}
\begin{document}

\title{Enhancing Code-switched Text-to-Speech Synthesis Capability in Large Language Models with only Monolingual Corpora}
\author{\IEEEauthorblockN{Jing Xu$^{\dagger}$, Daxin Tan$^{\star}$, Jiaqi Wang$^{\dagger}$, Xiao Chen$^{\star}$}
\IEEEauthorblockA{
\textit{$^{\dagger}$The Chinese University of Hong Kong}\\
\textit{$^{\star}$Noah's Ark Lab, Huawei} \\
jingxu.se@link.cuhk.edu.hk, jqwang23@cse.cuhk.edu.hk, \{tan.daxin1, chen.xiao2\}@huawei.com}}
\maketitle

\begin{abstract}
While Large Language Models (LLMs) have shown potential in speech generation and recognition, their applications are mainly confined to monolingual scenarios, with limited explorations in code-switched (CS) contexts. In this paper, we propose a Code-Switched Large Language Model (CS-LLM) to enhance the code-switched text-to-speech synthesis (CS TTS) capability in LLMs with only monolingual corpora. Specifically, we begin by enhancing the multilingual speech processing ability of LLMs through multilingual speech recognition and synthesis tasks. Then, we develop an effective code-switched (CS) data construction strategy that splits and concatenates words from different monolingual speech corpora to equip LLMs with improved CS TTS ability. Experiments show that our approach outperforms baselines in CS TTS in terms of naturalness, speaker consistency and similarity even with limited data. Additionally, the constructed CS data further improves multilingual speech synthesis and recognition.\footnote{Demo page: \href{https://jingxu96.github.io/csllm-demo/}{https://jingxu96.github.io/csllm-demo/}} 
\end{abstract}

\begin{IEEEkeywords}
Multilingual Text-to-Speech Synthesis, Code-switched Speech Synthesis,  Code-switched Data Construction.
\end{IEEEkeywords}

\section{Introduction}
\label{sec:intro}

Large Language Models (LLMs) have demonstrated remarkable capabilities in handling diverse natural language processing (NLP) tasks, such as text generation, translation, and summarization\cite{gpt3,gpt4}. 
Motivated by these achievements, recent studies have sought to equip LLMs with speech capabilities including recognition, synthesis, and understanding\cite{speechgpt,valle,salmonn,qwen2audio,ltu}. 
However, most of these efforts remain limited to monolingual scenarios, restricting their applications in multilingual contexts.


One prevalent phenomenon in multilingual communities is code-switching, where speakers alternate between two or more languages within a sentence\cite{csdefinition}. Code-switching poses significant challenges for speech technologies, particularly in code-switched text-to-speech synthesis (CS TTS), which aims to generate speech from code-switched text. This task faces unique difficulties in ensuring prosody consistency, naturalness and intelligibility across language boundaries. Moreover, the lack of high-quality code-switched data exacerbates these challenges, as collecting such data is both costly and time-consuming. 

Early attempts to address these issues involve adapting voices from an average voice\cite{csaverage,csvoice}, employing unit mapping techniques like phoneme and frame mapping\cite{csphone,csframe}, and utilizing bilingual Phonetic Posterior-Grams (PPGs)\cite{csbppg}. 
However, they often involve intricate model designs such as adversarial loss\cite{csbppg} and separate language encoders\cite{csencoder}. 
Recent advancements, such as VALL-E X\cite{vallex}, have explored code-switched text-to-speech, but still rely on frame-level language identification and complex architecture design, which do not fully resolve issues inherent in the CS TTS task. 

In this paper, we address the challenge of handling code-switched text by leveraging multilingual capabilities of LLMs, which excel in processing linguistic diversity. However, LLMs are primarily designed for text modality and lack the ability to generate or process multilingual speech for the CS TTS task. To address this, we integrate multilingual text-to-speech synthesis (TTS) and multilingual automatic speech recognition (ASR) tasks into LLMs. Inspired by the great success of self-supervised learning (SSL) models, such as Wav2vec 2.0\cite{wav2vec}, HuBERT\cite{hubert}, and WavLM\cite{wavlm}, we utilize discrete SSL representations, which capture richer semantic information\cite{layerwise,speechgpt}, to effectively align speech and text modalities. These discrete SSL representations are used as the speech input for ASR task and output for TTS task in CS-LLM, enhancing its multilingual speech processing capabilities.

Furthermore, we develop a novel and efficient code-switched (CS) data construction strategy using solely monolingual corpora to enhance CS TTS capability in CS-LLM. 
Existing CS data construction strategies\cite{jecs,dacs} are often complex, typically relying on word insertion and translation. Besides, they often require additional speech synthesis systems for data construction. 
For example, \cite{jecs} requires bilingual speakers to train two separate monolingual TTS systems, while \cite{dacs} involves building a CS TTS system in advance to generate CS data for CS speech recognition. Thus, these strategies are often cumbersome, difficult to scale, and less effective. In contrast, our approach simplifies this process by just splitting and concatenating words from different monolingual speech corpora, eliminating the need for additional speech synthesis systems.

\begin{figure*}
\centering
\begin{minipage}[b]{0.496\linewidth}
  \centering
  \centerline{\includegraphics[width=8.8cm]{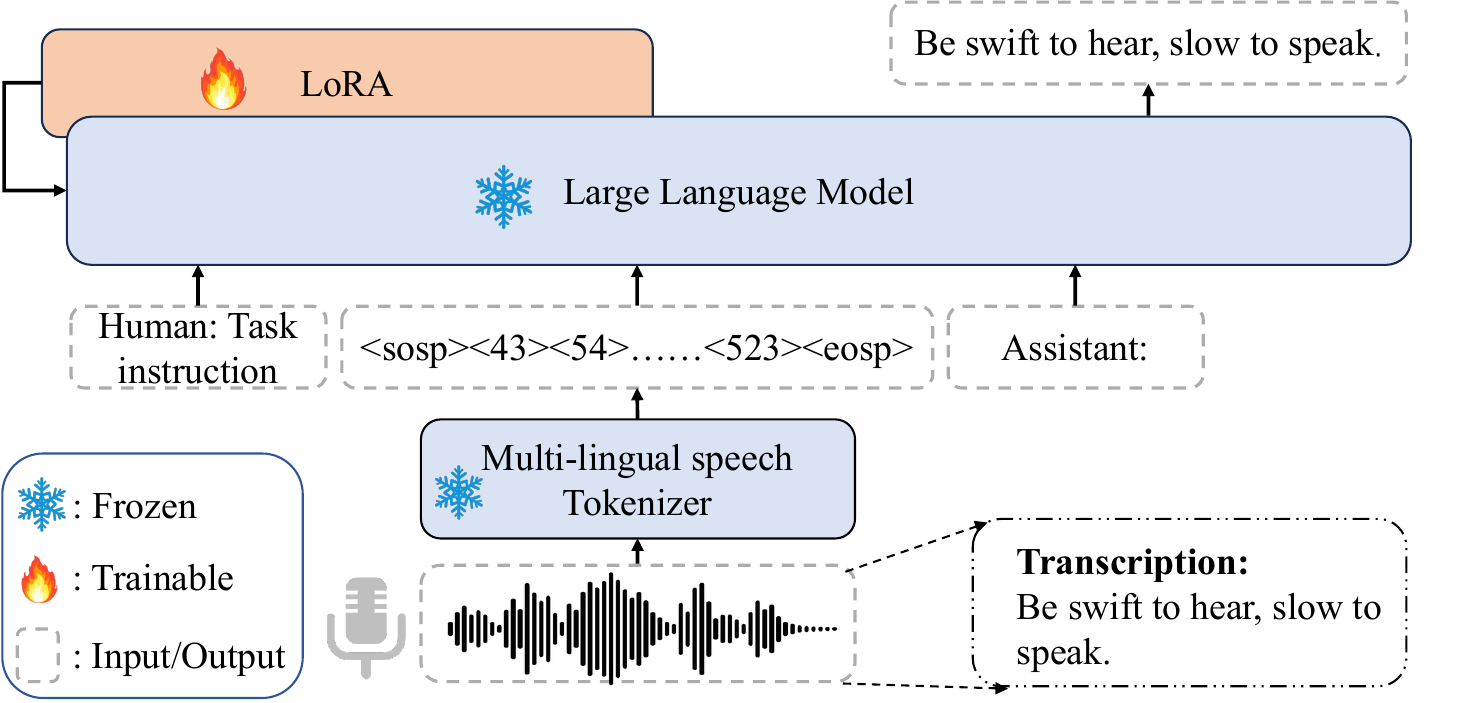}}
  \centerline{(a) ASR task}\medskip
\end{minipage}
\begin{minipage}[b]{0.496\linewidth}
  \centering
  \centerline{\includegraphics[width=8.8cm]{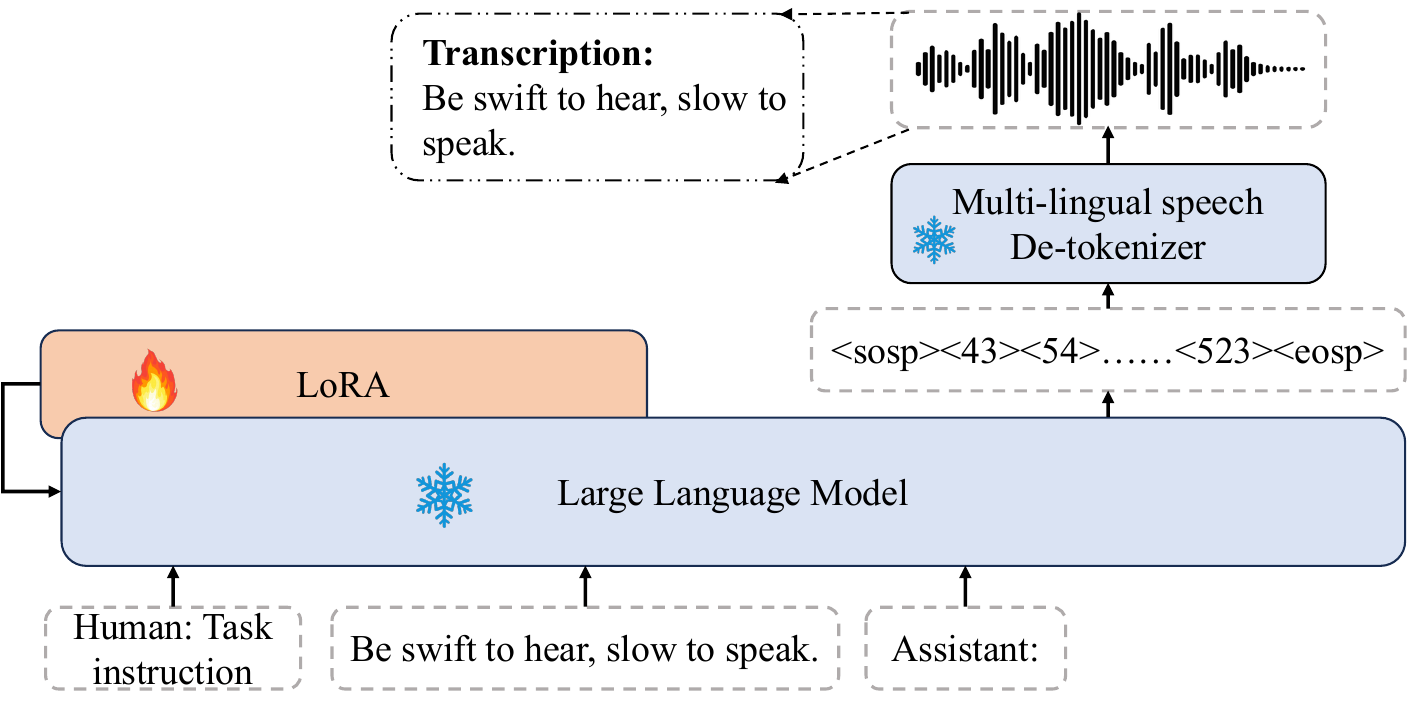}}
  \centerline{(b) TTS task}\medskip
\end{minipage}
\hfill
%
\caption{A pipeline of CS-LLM when implementing multilingual ASR task (Figure (a)) and multilingual TTS task (Figure (b)). Note parameters of all adapters and the LLM backbone are the same when implementing different tasks. }
\label{fig:CS-LLM struct}

\end{figure*}

In summary, we introduce a Code-switched Large Language Model (CS-LLM) to enhance code-switched speech synthesis ability in LLMs using only monolingual corpora. We first improve multilingual speech processing capabilities in LLMs and then propose a novel and efficient CS data construction strategy to enhance CS TTS. 
Extensive experiments show the effectiveness of our proposed methods. Our contributions include:
 \begin{itemize}
    \item We enhance the multilingual speech processing capabilities of CS-LLM through multilingual TTS and ASR tasks.
     \item We develop a novel and efficient CS data construction strategy that splits and concatenates words from different monolingual speech corpora, significantly reducing the dependency on high-quality CS data for the CS TTS task.
    \item Both subjective and objective experimental results demonstrate that our CS-LLM model outperforms baselines in CS TTS in terms of naturalness, speaker consistency and similarity. Furthermore, the constructed CS data further improves multilingual ASR and TTS performance of LLMs.
 \end{itemize}
\section{Related works}

\subsection{Speech language models}

The development of Speech Large Language Models (SpeechLLMs) has accelerated in recent years, with significant advancements in speech generation, recognition, and multi-modal integration. Early works like GSLM\cite{gslm} laid the foundation for speech-based LLMs, followed by models such as AudioLM\cite{audiolm}, which introduced semantic and acoustic tokens to capture both linguistic and audio characteristics. These models adopt a multi-stage generation process (semantic → coarse acoustic → fine-grained acoustic) to ensure semantic coherence and accurate speech synthesis.

Building upon these methods, VALL-E\cite{valle} reformulated text-to-speech (TTS) as a conditional language modeling problem, using neural codec tokens to represent speech. While effective, its multi-stage pipeline—including an autoregressive (AR) component followed by a non-autoregressive (NAR) residual model—introduces complexity in training and inference. Subsequent extensions like VALL-E X\cite{vallex} enable cross-lingual synthesis, while Spear-TTS\cite{speartts} incorporates multiple AR models to enhance multi-speaker synthesis with minimal supervision.

Beyond TTS, recent SpeechLLMs aim to bridge speech and text modalities. SpeechGPT\cite{speechgpt} fine-tunes models on cross-modal tasks (ASR, TTS) and chain-of-modality question answering (QA), integrating speech and text-based responses. Similarly, Spectron\cite{spectron} employs text as a bridge for spoken QA, utilizing spectrogram representations rather than discrete tokens. 

Newer multimodal architectures, such as AudioPALM\cite{audiopalm} and VioLA\cite{viola}, focus on joint speech-text training for ASR, TTS, and speech translation. Further developments by VoxtLM\cite{voxtlm} and SUTLM\cite{SUTLM} introduce joint speech-text LMs, enhancing speech/text continuation tasks. Our work leverages speech-text training for multilingual ASR and TTS tasks to enhance multilingual speech understanding and generation capabilities of large language models, facilitating the further integration of code-switched speech synthesis.

\subsection{Disentanglement leveraging Self-supervised model}
Self-supervised learning (SSL) has significantly advanced speech representation learning by enabling models to extract rich linguistic, prosodic, and speaker-specific features without labeled data. Early SSL frameworks, such as wav2vec 2.0\cite{wav2vec} and HuBERT\cite{hubert}, leveraged contrastive and masked prediction objectives to learn robust speech embeddings. w2v-BERT\cite{w2v-bert} further enhanced SSL by combining wav2vec-style continuous features with discrete tokenization, improving both ASR and generative speech tasks. Recent models, such as Whisper\cite{whisper} and USM\cite{usm}, demonstrate the effectiveness of multi-task SSL, where speech models trained on diverse speech corpora learn generalizable features applicable to speech recognition, translation, and synthesis. 

Representations learned by these SSL models capture rich content (phonetic, semantic) and style (speaker identity, emotion, prosody) information, allowing models to separate speaker-specific information from linguistic content. Thus, SSL representations are suitable for tasks requiring disentanglement such as voice conversion (VC), expressive TTS, and zero-shot TTS. 

Traditional VC approaches rely on parallel corpora and supervised learning, but SSL-based methods reduce data dependency by leveraging pre-trained SSL representations. VITS\cite{vits} utilize SSL speech representations to disentangle content from speaker identity, allowing non-parallel conversion. AutoVC\cite{autovc} first introduced a bottleneck-based approach for speaker-independent content extraction, later improved by AdaIN-VC\cite{adainvc} using adaptive instance normalization.

Beyond continuous speech features, discrete SSL representations have been widely applied, enabling models to learn robust speech units for semantic understanding and disentanglement. SpeechGPT\cite{speechgpt} applies K-means clustering to HuBERT representations, generating semantic tokens that align more effectively with text. Similarly, SpeechTokenizer\cite{speechtokenizer} employs residual vector quantization to decompose speech representations into distinct content and acoustic tokens, facilitating more robust speech codecs for speech synthesis.

In this work, we employ K-means clustering to discretize representations extracted from an adapted multilingual HuBERT, generating discrete speech tokens that effectively remove speaker-specific attributes while preserving both linguistic and semantic information. This disentanglement property enhances the ability of large language models (LLMs) to interpret discrete speech tokens across different speakers and languages, thereby enabling the construction of code-switched speech samples using only monolingual corpora.

\section{Methodology}
\label{sec:method}
In this section, we first introduce the components of our proposed CS-LLM model (Subsection~\ref{CS-LLM}).  Next, we illustrate the CS data construction strategy using only monolingual corpora (Subsection~\ref{data}). 
Finally, we put forward two different training strategies to enhance the CS TTS capability of LLMs via monolingual corpora and the constructed CS data (Subsection \ref{training}). 
\subsection{Model structure}\label{CS-LLM}
As illustrated in Figure~\ref{fig:CS-LLM struct}, CS-LLM consists of three primary components: a multilingual speech tokenizer, a large language model (LLM) backbone, and a multilingual speech de-tokenizer. These components work together to enhance multilingual speech processing capabilities in LLMs. 
\subsubsection{Multilingual speech tokenizer}
We employ the multilingual speech tokenizer in \cite{seamless} to convert multilingual speech waveforms into discrete units. The multilingual speech tokenizer consists of an adapted HuBERT model and a K-means model. The adapted HuBERT model enhances the Mandarin capabilities of initial HuBERT model without compromising English performance through designed adaptation and preservation strategies. Subsequently, the K-means model discretizes multilingual speech representations, extracted by the adapted HuBERT model, into cluster indices. These de-duplicated cluster indices are used as discrete speech units, serving as the input for ASR task. 
\subsubsection{LLM backbone}
We use LLaMA 3 8B\cite{llama3.2} as the LLM backbone, a decoder-only model pre-trained on more than 15 trillion tokens. Importantly, over 5\% of the LLaMA 3 pre-training dataset comprises of high-quality non-English data from over 30 languages, ensuring strong textual multilingual capabilities. 

To make CS-LLM capable of processing discrete speech units, we expand the initial vocabulary and the corresponding embedding matrix of LLaMA 3. Assuming the original vocabulary $V$ with size $|K|$, we expand it with an additional set of discrete speech units $V'$ with size $|K'|$. The final expanded vocabulary $V''$ with size $|K''|$ is the union of the original vocabulary and the new set of discrete speech units:
$$V''=V\cup V'$$
The original word embedding matrix can be denoted as $E\in \mathbb{R}^{|K|\times d}$, where the dimension of each word embedding equals to $d$.
To accommodate the expanded vocabulary, we create a new word embedding matrix $E'\in \mathbb{R}^{|K''|\times d}$, where the first $|K|$ rows of $E'$ are copied from the original word embedding matrix $E$ to preserve the original word embeddings. The remaining rows are randomly initialized:
$$E'[0:|K|,:]=E$$
This expansion allows CS-LLM to more effectively process both speech and text data, thereby enhancing its multilingual speech processing capability during training on multilingual TTS and ASR tasks.

As shown in Figure~\ref{fig:CS-LLM struct}, CS-LLM receives task instructions, task-related inputs, and role information. The task instructions for each task are shown in Table~\ref{tab: task}. For the ASR task, task-related input is discrete speech units generated by the multilingual speech tokenizer, and the output is the corresponding text. For the TTS task, the task-related input is text and the output is a sequence of discrete speech units. To efficiently finetune LLMs, we employ Low Rank Adaptation (LoRA)\cite{lora}.
\begin{table}[ht]
\caption{Illustration of task instruction}
\label{tab: task}
\makebox[\linewidth][c]{
    \resizebox{\linewidth}{!}{
    \begin{tabular}{ccc}
    \hline
     & EN & ZH     \\ \hline
   TTS & Please speak the sentence. & \begin{CJK}{UTF8}{gbsn}请说出下面的句子。\end{CJK} \\
   ASR&Please transcribe the speech. &\begin{CJK}{UTF8}{gbsn}请把语音转录成文本。\end{CJK} \\ \hline
    \end{tabular}
    }
}
\end{table}
\subsubsection{Multilingual speech de-tokenizer}

\begin{figure}[h]
\centering
\includegraphics[width=\linewidth]{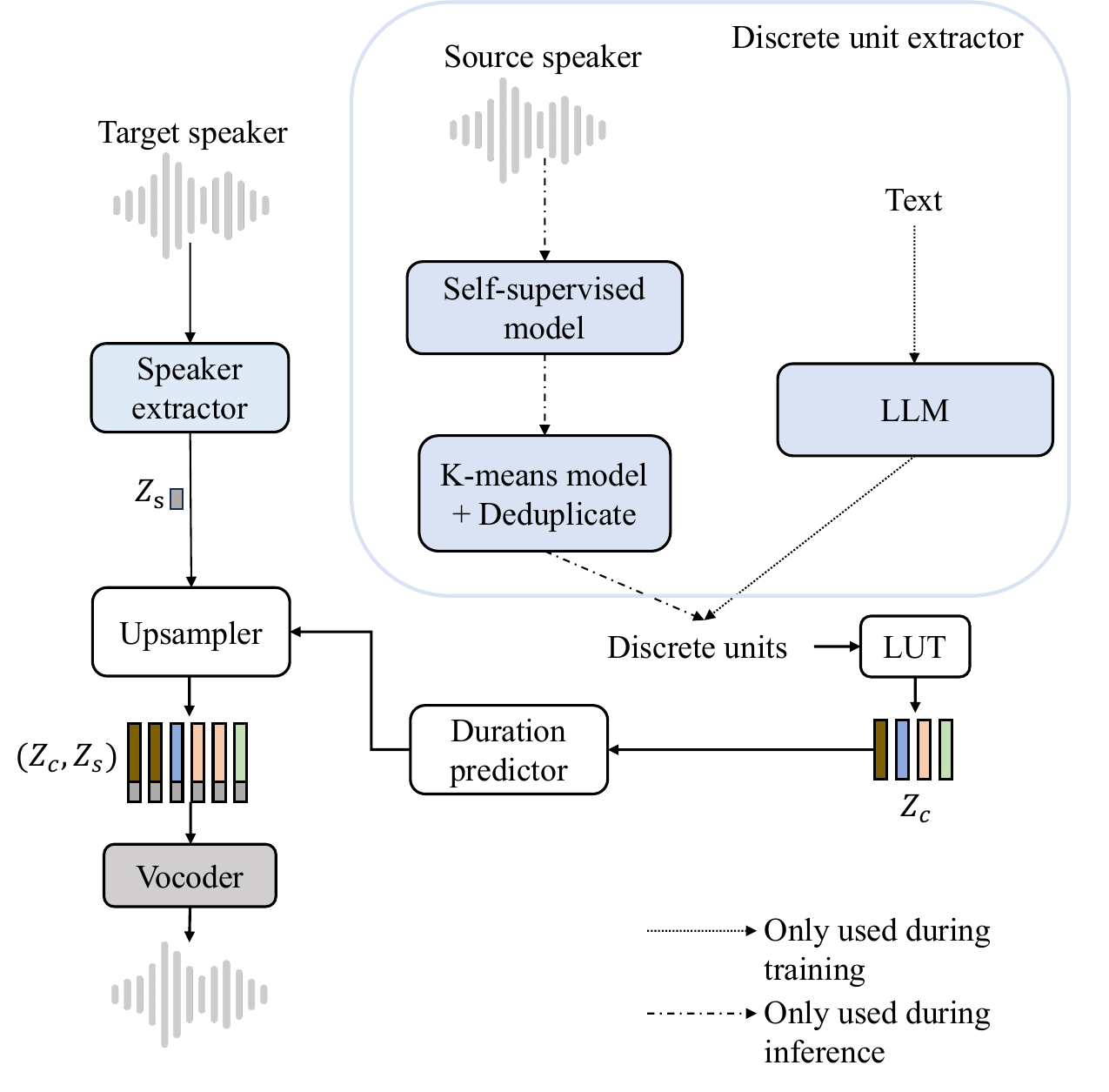}
%
\caption{The structure of multilingual speech de-tokenizer. }
\label{fig:detokenizer}

\end{figure}

The multilingual speech de-tokenizer converts discrete speech units generated by the LLM into multilingual speech waveform when conducting TTS task. As shown in Figure~\ref{fig:detokenizer}, the multilingual speech de-tokenizer is comprised of a discrete unit extractor, a speaker extractor and a neural vocoder. 
\begin{itemize}
    \item \textbf{Discrete unit extractor.} During training, discrete unit extractor aims to discretize representations extracted by self-supervised model into cluster indices using K-means model. The sequence of cluster indices are de-duplicated and serve as discrete units. During inference, the discrete units are generated by LLM based on the given text. 
    \item \textbf{Speaker extractor.} To enable multi-speaker speech synthesis, speaker embeddings $Z_s$ are extracted from WavLM-SV\cite{wavlm}, a model fine-tuned on the speaker verification task. 
    \item \textbf{Neural vocoder}. A neural vocoder is employed to transform speech discrete units back into speech waveform conditioned on the speaker embedding $Z_s$. As shown in Figure~\ref{fig:detokenizer}, first the discrete units are converted into continuous representations $Z_c$ via look-up tables (LUT). The sequences $Z_c$ and $Z_s$ are then up-sampled based on the duration predicted by the duration predictor. Speaker embedding $Z_s$ is concatenated to each frame of up-sampled embedding $Z_c$. Then, a vocoder, whose architecture is similar to Hifi-GAN, is trained to transform the concatenated sequence into waveform. 
\end{itemize}

\subsection{Code-switched Data construction strategy}\label{data}
\begin{figure}[t]
\centering
\includegraphics[width=\linewidth]{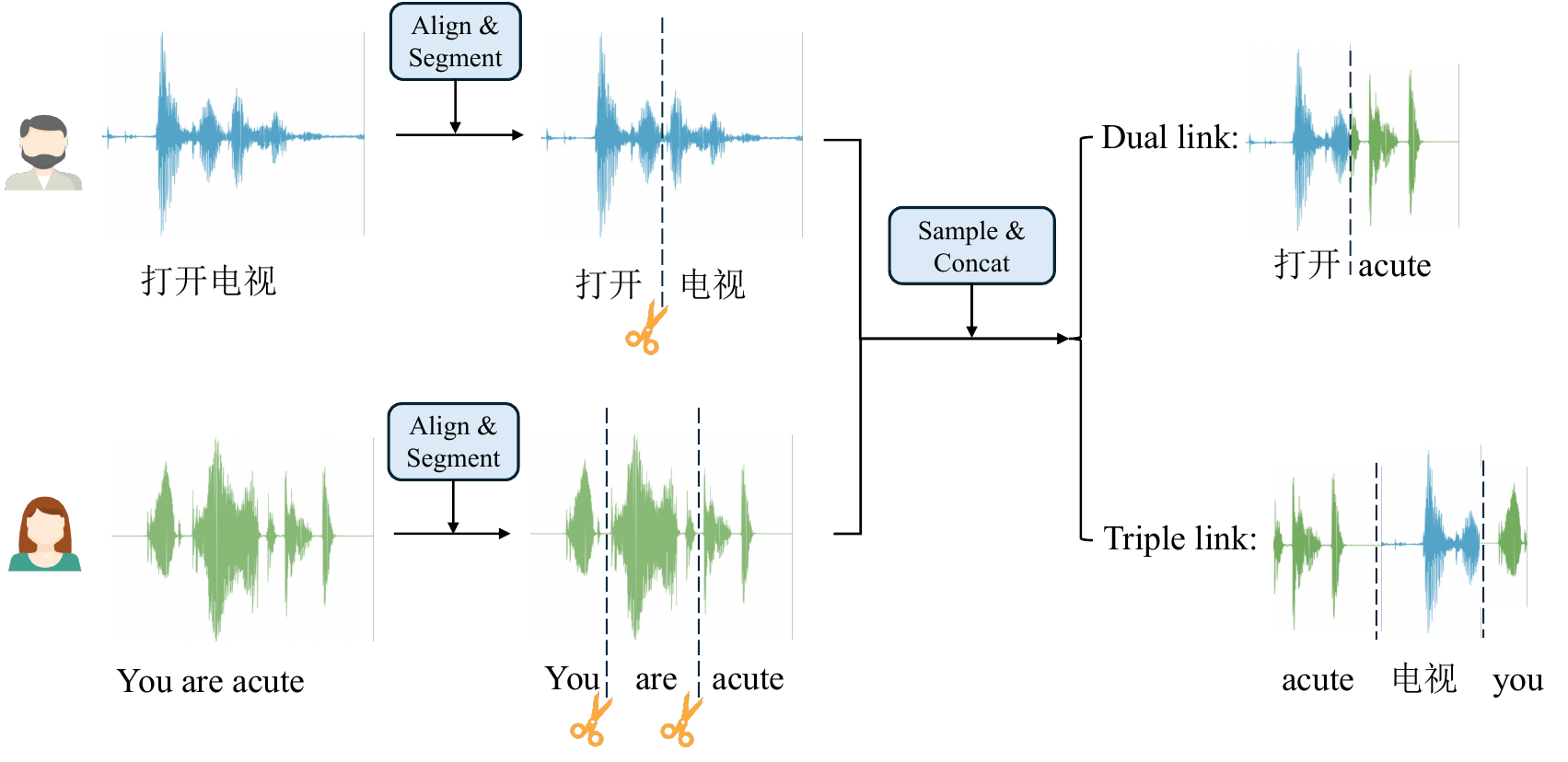}
%
\caption{The illustration of CS data construction strategy. }
\label{fig:data construct}
\end{figure}

The discrete speech units extracted from SSL models capture more phonetic information\cite{analysisphone}, while discarding speaker-related information. This motivates us to explore whether we can reduce the reliance on high-quality CS data by constructing CS sentences through simple concatenation of words from different languages. Since these discrete units retain semantic and phonetic information and LLMs possess strong in-context learning capabilities, we hypothesize that LLMs can effectively learn the patterns of CS speech even from concatenated data. To this end, we propose a two-step CS data construction strategy, as shown in Figure~\ref{fig:data construct}:

\begin{itemize}
    \item Align and Segment: We first align text with the corresponding speech from monolingual corpora to determine the duration of each Mandarin character and English word. Mandarin sentences are then segmented into words, and the corresponding duration for each word is calculated. We then segment Mandarin and English sentences based on word-level durations.
    \item Sample and concatenate: We randomly select one word from each language and retrieve the corresponding speech clip. These clips, originating from different languages and speakers, are concatenated to form a CS sentence.
    
\end{itemize}
We define three types of constructed CS dataset based on sentence structure and dataset composition:
\begin{itemize}
    \item Dual-link: All sentences are formatted as Lang1-Lang2, with each language clip having an equal probability (0.5) of appearing at begining of the sentence. 
    \item Triple-link: All sentences are formatted as Lang1-Lang2-Lang1, where the first and last clips are in the same language, and the middle clip comes from a different language. Each language has an equal chance of appearing at the beginning.
    \item Mixed: The dataset includes an equal number of Lang1-Lang2 and Lang1-Lang2-Lang1 formatted sentences.
\end{itemize}
\begin{table*}
\setlength \tabcolsep{2pt}
\centering
\caption{Objective and subjective evaluations of different models. In the bracket of our proposed systems, the term before dash denotes data construction strategy (with 'no' indicating no data construction) and the term after dash represents training hours per language.}
\label{tab: overall}
\resizebox{\linewidth}{!}{
\begin{tabular}{lccccccccccccc}
\hline
\multirow{2}{*}{System}   & \multirow{2}{*}{Tasks}  & \multicolumn{2}{c}{ASR(WER$\downarrow$)}            & \multicolumn{2}{c}{TTS(WER$\downarrow$)}  & \multicolumn{2}{c}{TTS(MOS$\uparrow$)} & \multicolumn{3}{c}{CS TTS}\\ 
\cline{3-11}
  &                                   & librispeech & aishell2 &librispeech & aishell2 & librispeech & aishell2      &  MOS$\uparrow$ &  SCS-EN$\uparrow$ &  SCS-ZH$\uparrow$\\ 
\hline\hline
\multicolumn{4}{l}{\textbf{Ground truth}} \\
\hspace{0.4em}GT    & -   &  2.03\%   & 2.83\% & -   &   -   &   4.591$\pm$0.221      &   3.621$\pm$0.431    &   - &- &-   \\ \hline\hline
\multicolumn{4}{l}{\textbf{Baselines}} \\
\hspace{0.4em}SpeechGPT     & ASR, TTS      &  16.94\%   & - & 40.87\%   &     -  &    3.838$\pm$0.356     &    -  &   - &- &-   \\
\hspace{0.4em}Initial Hubert   & ASR, TTS   &  27.87\%   & - & 7.70\%   &   -   &     \textbf{4.566$\pm$0.250}    &     -  &   - &- &-   \\
\hspace{0.4em}VALL-E X       & TTS, CS TTS   &   -  & - &  31.59\%  &   19.52\%    &   3.956$\pm$0.400    & 3.360$\pm$0.429&  2.900$\pm$0.394 &    0.842   &     0.909 \\
 \hline\hline
\multicolumn{10}{l}{\textbf{Our proposed methods}} \\
\hspace{0.4em}CS-LLM(no-500)   & ASR, TTS & 14.18\%  &  12.15\% & 8.46\%  &  9.14\%     & 4.319$\pm$0.285& 4.132$\pm$0.285 & 2.456$\pm$0.293  & 0.829 &0.913  \\
\hspace{0.4em}CS-LLM(mixed-500)  & ASR, TTS, CS TTS & 13.62\%  &  12.62\% & 10.65\%  &  9.06\%  & 4.281$\pm$0.281  & 4.263$\pm$0.272   & 3.875$\pm$0.323 & 0.842 &0.910  \\
\hspace{0.4em}CS-LLM(no-1000)  & ASR, TTS & 10.46\%  &  9.00\% & 7.63\%  &  10.22\%  &  4.244$\pm$0.313 &  4.110$\pm$0.304 & 1.7625$\pm$0.389 & \textbf{0.868}&0.901  \\
\hspace{0.4em}CS-LLM (mixed-1000) & ASR, TTS, CS TTS &  \textbf{9.11\%}  &  \textbf{7.89\%} & \textbf{6.96\%}  & \textbf{9.01\%}  & 4.263$\pm$0.304  & \textbf{4.447$\pm$0.237}  & \textbf{4.050$\pm$0.306} &  0.845  & \textbf{0.918}\\ \hline

\hline
\end{tabular}
}
\end{table*}
\subsection{Training strategy}\label{training}
We propose two training strategies for enhancing the CS TTS capability of CS-LLM, depending on when the constructed CS data is incorporated:
\begin{itemize}
    \item One-stage training strategy: In this strategy, the constructed CS data is combined with two monolingual corpora, and the CS-LLM model is trained to perform multilingual ASR, multilingual TTS, and CS TTS tasks simultaneously.
    \item Two-stage training strategy: In this strategy, CS-LLM is first trained to perform multilingual ASR and multilingual TTS tasks using monolingual corpora. The resulting model is then further finetuned with the constructed CS data for the CS TTS task, with only limited parameters updated.
\end{itemize}

\section{Implementation}
\label{sec:setup}

We use Montreal Forced Alignment (MFA)\footnote{https://montreal-forced-aligner.readthedocs.io/en/latest/index.html} to align speech-text pairs and the open-sourced tool jieba\footnote{https://github.com/fxsjy/jieba} to segment Mandarin sentences. To construct CS dataset, we randomly sample one word from each language to form the code-switched sentences. 

For the multilingual speech tokenizer and de-tokenizer, we employ the adapted mHuBERT model, K-means model and unit-HiFiGAN from \cite{seamless}. The mHuBERT model is adapted to handle both Mandarin and English speech with a 10:1 adaptation ratio. The K-means model, with 1000 clusters, is obtained by clustering the English and Mandarin representations using the MiniBatchKMeans algorithm. We leverage WavLM\footnote{https://huggingface.co/microsoft/wavlm-base-sv} to extract speaker embedding in multilingual speech de-tokenizer.

We adopt LLaMA 3-8B\cite{llama3.2} as the LLM backbone in all experiments, incorporating the LoRA\cite{lora} adapter. The input and output of the TTS task are reversed to adapt for the ASR task. For the CS TTS task, the task instruction is "Please speak the code-switched sentence." The constructed CS sentences are formatted for both CS TTS and CS ASR tasks.  All CS-LLM models are trained for 2 epochs with a batch size of 4. The LoRA rank is set to 1024 unless otherwise specified.

\section{Experimental results}
\label{sec:results}
\subsection{Datasets}
For training, we use AISHELL-2\cite{aishell2} and LibriSpeech\cite{librispeech} as the Mandarin and English corpora, respectively. AISHELL-2 consists of 1000 hours of clean Mandarin reading speech, while LibriSpeech contains 960 hours of English reading speech. All speech samples are uniformly down-sampled to \SI{16}{\kilo\hertz}. Unless otherwise specified, the constructed CS dataset comprises 10 hours of constructed CS speech. For the CS TTS inference, we randomly sample 1000 samples from ASRU-2019 Challenge test set, using the text as input.
\subsection{Baselines}
We compare CS-LLM against the following three baselines:
\begin{itemize}

    \item \textbf{VALL-E X}:  It takes phoneme sequences of source and target text, along with source acoustic tokens as prompts, to produce target acoustic tokens. We utilize the open-sourced model trained on 704 hours and 598 hours of English and Mandarin TTS data, respectively.
    \item \textbf{SpeechGPT}: It uses discrete units from mHuBERT and a LLaMA 2 7B model as the backbone. It is trained for English ASR and TTS tasks via cross-modal instruction fine-tuning.
    \item \textbf{Initial HuBERT}: For a fair comparison, we train another model with units from mHubert and the LLaMA 3 8B backbone using 500 hours of English speech for 2 epochs.
\end{itemize}

\subsection{Evaluation metrics}
We evaluate the performance of the CS TTS task with a 5-scale Mean Opinion Score (MOS) test (1-bad, 2-poor, 3-fair, 4-good, 5-excellent) and Speaker Cosine Similarity (SCS). We separately select 10 random speakers from TIMIT\cite{timit} and ASCEND\cite{ascend} datasets, using only "zh" labeled sentences from ASCEND. SCS is computed between generated speech and speech from unseen speaker using WavLM\cite{wavlm}. These speakers are not included during the training of CS-LLM, multilingual speech tokenizer and de-tokenizer. We assess Word Error Rate (WER) for both ASR and TTS tasks and conduct a 5-scale MOS test for the TTS task. In this paper, for Mandarin, WER refers to Character Error Rate (CER). Transcripts of generated English and Mandarin speech are obtained from Whisper large-v3\cite{whisper} and Paraformer-ZH\cite{paraformer} respectively.  For all MOS tests, 20 listeners are asked to assess 10 random samples from each system based on naturalness, with MOS scores computed at 95\% confidence intervals. 
\subsection{Main results}

 As shown in Table~\ref{tab: overall}, the proposed CS data construction strategy enhances the CS TTS capability of CS-LLM, improving speech naturalness (MOS scores increases from 2.456 to 3.875). Besides, it outperforms VALL-E X in naturalness (4.263 vs. 3.360) and achieves comparable speaker consistency (0.842 vs. 0.842) and similarity (0.910 vs. 0.909) at a lower data scale. As the data scale increases, CS-LLM surpasses VALL-E X in all metrics. The constructed CS data even boosts ASR and TTS performance, showing the potential of our CS data construction strategy. Specifically, CS-LLM with a mixed data construction strategy outperforms all three baselines in multilingual ASR and TTS tasks objectively, achieving relative WER reductions of up to 67.31\% for English ASR and 82.97\% for English TTS. While our models slightly underperform initial HuBERT in English subjectively, it is likely due to the single-speaker synthesis. Notably, CS-LLM performs significantly better with 1000 training hours per language, which will be the standard setting for subsequent experiments if not specified.

\subsection{Data construction and training strategies comparison}

We also evaluate the impact of different data construction and training strategies. In Table \ref{tab: data construciton}, the bracketed item represents the data construction or training strategy, where “mixed ttS” refers to a setting where sentences are formatted using the mixed data strategy, and the training samples consist exclusively of TTS data. We can see that the mixed data construction strategy achieves the best performance across all tasks possibly due to the diversity in data format. Including the constructed ASR format CS data further contributes to improved performance. In the two stage training strategy, LoRA with a rank of 8 is added to the linear layers in the Transformer, and the model is trained for 2000 steps. Despite not updating the embedding layer or LM head in the second stage, the CS-LLM effectively performs the CS TTS task, with minimal impacts on other tasks. The CS TTS task instruction, "Please speak the code-switched sentence", was not seen during the first training stage. This highlights the strong in-context learning capabilities of LLMs and demonstrates the effectiveness of guiding them through multilingual ASR and TTS tasks for implementing code-switched text-to-speech synthesis task. 
\begin{table}[ht]
\setlength \tabcolsep{2pt}
\centering
\caption{Comparison of data construction and training strategy}
\label{tab: data construciton}
\resizebox{\linewidth}{!}{
\begin{tabular}{lccccccccc}
\hline
\multicolumn{1}{c}{\multirow{2}{*}{System}}        & \multicolumn{2}{c}{ASR (WER$\downarrow$)}      & \multicolumn{2}{c}{TTS (WER$\downarrow$)}     &\multicolumn{2}{c}{CS TTS (SCS$\uparrow$)} \\ 
\cline{2-7}
                                        & librispeech &  aishell2 & librispeech & aishell2 & EN & ZH \\ 
\hline
\multicolumn{5}{l}{\textbf{Different data construction}} \\ 
\hspace{0.4em}CS-LLM(mixed)         &   \textbf{9.11}\%        &     \textbf{7.89\%}   &   \textbf{6.96\%}      & 9.01\% & \textbf{0.845}& \textbf{0.918} \\
\hspace{0.4em}CS-LLM(triple)   &        14.71\%    &    10.09\%   &     7.87\%      &   8.90\% &  0.842  &0.913\\
\hspace{0.4em}CS-LLM(dual)  &         9.25\%    &      9.35\%    &    7.38\%          & 8.93\% &   0.832 &0.897\\ 
\hspace{0.4em}CS-LLM(mixed tts) &    9.65\%        &    8.11\%      &       7.91\%       & 8.68\%&   0.840 &0.913 \\ 
\hline\hline
\multicolumn{5}{l}{\textbf{Different training strategies}} \\ 
\hspace{0.4em}CS-LLM(2 stage)&    10.71\%        &    7.97\%      &       8.74\%       & \textbf{8.40\%} & 0.838   & 0.912\\ 
\hline
\end{tabular}

}
\end{table}
\subsection{Ablation study}

We perform an ablation study on training hours per language, LoRA rank, and constructed CS data size to assess the effectiveness of our proposed CS-LLM system and data construction strategy. As shown in Table ~\ref{tab: training hour},  increasing training hours yields better speech processing capability of CS-LLM. As shown in Table ~\ref{tab: Lora} and Table~\ref{tab: training hour}, while adding LoRA rank also contributes, its effect is less significant compared to increasing training hours. Table ~\ref{tab: constructed} shows that more constructed CS data does not always improve ASR and TTS performance, possibly due to the different data distribution compared to traditional ASR and TTS datasets. Excessive constructed data may lead to overfitting. Notably, even with just 1 hour of constructed data, performance improvements can be observed across almost all tasks, demonstrating the efficiency of our CS data construction strategy.

\begin{table}[ht]
\centering
\caption{Ablation study on training hours per language}

\label{tab: training hour}
\resizebox{\linewidth}{!}{
\begin{tabular}{lccccccccc}
\hline
\multicolumn{1}{c}{\multirow{2}{*}{System}}        & \multicolumn{2}{c}{ASR (WER$\downarrow$)}      & \multicolumn{2}{c}{TTS (WER$\downarrow$)}        \\ 
\cline{2-5}
                                        & librispeech &  aishell2 & librispeech & aishell2  \\ 
\hline
dur=100         &   20.55\%        &     17.61\%   &   11.58\%      & 17.26\% \\
dur=360   &        16.17\%    &    18.47\%   &     8.14\%      &   10.06\%\\
dur=500  &      13.62\%  &  12.62\% & 10.65\%  &  9.06\% \\ 
dur=1000   &    \textbf{9.11\%}        &    \textbf{7.89\%}      &       \textbf{6.96\%}       & \textbf{9.01\%}\\ 
\hline
\end{tabular}
}

\end{table}
\begin{table}[ht]
\centering
\caption{Ablation study on LoRA rank}

\label{tab: Lora}
\resizebox{\linewidth}{!}{
\begin{tabular}{lccccccccc}
\hline
\multicolumn{1}{c}{\multirow{2}{*}{System}}        & \multicolumn{2}{c}{ASR (WER$\downarrow$)}      & \multicolumn{2}{c}{TTS (WER$\downarrow$)}        \\ 
\cline{2-5}
                                        & librispeech &  aishell2 & librispeech & aishell2  \\ 
\hline
\multicolumn{5}{l}{\textbf{without data construction strategy}} \\ 
\hspace{0.2em}CS-LLM (r=1024)   &       10.46\%       &       9.00\%   &        7.63\%      &    10.22\%\\
\hspace{0.2em}CS-LLM (r=512)   &        11.45\%      &     10.00\%     &      7.57\%       &  9.03\% \\
\hspace{0.2em}CS-LLM (r=256)          &     11.42\%        &    11.33\%       &     8.36\%        & 9.53\% \\

 \hline
\multicolumn{5}{l}{\textbf{with mixed data construction strategy}} \\
\hspace{0.2em}CS-LLM (r=1024)   &       \textbf{9.11\%}       &      \textbf{7.89\%}     &       \textbf{6.96\%}       &   \textbf{9.01\%}\\
\hspace{0.2em}CS-LLM (r=512)   &       10.27\%      &    8.82\%      &      7.26\%        &  9.04\% \\
\hspace{0.2em}CS-LLM (r=256)          &    13.73\%          &    9.86\%       &      7.12\%          &  11.07\%\\


\hline
\end{tabular}
}

\end{table}

\begin{table}[!h]
\centering
\caption{Ablation study on size of constructed CS data}
\label{tab: constructed}
\resizebox{\linewidth}{!}{
\begin{tabular}{lccccc}
\hline
\multirow{2}{*}{System}        & \multicolumn{2}{c}{ASR (WER$\downarrow$)}      & \multicolumn{2}{c}{TTS (WER$\downarrow$)}   \\ 
\cline{2-5}
                                    & librispeech &  aishell2 & librispeech & aishell2  \\ 
\hline
dur=1  &      9.66\%      &    8.47\%   &       7.90\%    & 9.74\%   \\
dur=10  &       \textbf{9.11\%}     &    \textbf{7.89\%}   &       6.96\%    & \textbf{9.01\%} \\
dur=20  &        9.80\%    &    9.94\%   &       \textbf{6.83\%}    &  9.36\%  \\
dur=50 &       11.64\%      &   8.03\%       &      7.17\%        &  9.38\% \\ 
\hline
\end{tabular}
}
\end{table}

\section{Conclusion}
\label{sec:typestyle}

In this paper, we propose CS-LLM to enhance the CS TTS capability in LLMs using solely monolingual corpora. Specifically, we begin by enhancing the multilingual speech processing capabilities in LLMs through multilingual speech synthesis and recognition tasks. Besides, we develop an effective CS data construction method which simply splits and concatenates words from different monolinugal speech corpora to enhance CS TTS. Experiments show that our CS data construction strategy enables CS-LLM to outperform baselines in CS TTS task and further enhances the performance of multilingual ASR and TTS tasks. 
Currently, we only focus on two languages in our system. In the future, we plan to incorporate more language pairs into our system. 

\newpage

\bibliographystyle{IEEEtran}
\bibliography{mybib}

\end{document}